\documentclass{aa}  

\usepackage{graphicx}
\usepackage{txfonts}
\usepackage{epsfig}
\usepackage{longtable}
\usepackage{amssymb} 
\newcommand{\sw}{$Swift$}

\def \inte {\emph{INTEGRAL}}

\def \sw {{\it Swift}}

\def \hcm {\hbox {\ifmmode $ atom cm$^{-2}\else atom cm$^{-2}$\fi}}

\def \apjl {ApJL}
\def \apjs {ApJS}

\def \aap {A\&A}

\begin{document}

   \title{Expected number of supergiant fast X-ray transients\\ in the Milky Way}

   \author{L. Ducci
          \inst{1,2}
          \and
          V. Doroshenko\inst{1}
          \and
          P. Romano\inst{3}
          \and
          A. Santangelo\inst{1}
          \and
          M. Sasaki\inst{1}
          }

   \institute{Institut f\"ur Astronomie und Astrophysik, Eberhard Karls Universit\"at, 
              Sand 1, 72076 T\"ubingen, Germany\\
              \email{ducci@astro.uni-tuebingen.de}
              \and
              ISDC Data Center for Astrophysics, Universit\'e de Gen\`eve, 16 chemin d'\'Ecogia, 1290 Versoix, Switzerland
              \and
              INAF, Istituto di Astrofisica Spaziale e Fisica Cosmica - Palermo,
              Via U.\ La Malfa 153, I-90146 Palermo, Italy
             }

   \date{}

 
  \abstract
  {In the past fifteen years a new generation of X-ray satellites led to the discovery
   of a subclass of high-mass X-ray binaries (HMXBs) with supergiant companions and a peculiar
   transient behaviour: supergiant fast X-ray transients (SFXTs).
   We calculate the expected number of Galactic SFXTs for the first time, using two different
   statistical approaches and two sets of data based on \sw\ and \inte\ surveys,
   with the aim to determine how common the SFXT phenomenon really is.
   We find that the expected number of SFXTs in the Galaxy is $\approx 37{+53\atop-22}$,
   which shows that SFXTs constitute a large portion of X-ray binaries with supergiant companions 
   in the Galaxy.
   We compare our estimate with the expected number of Galactic HMXBs 
   predicted from observations and evolutionary models and discuss the implications
   for the nature of SFXTs.}


   \maketitle
%

\section{Introduction}
\label{sect. intro}

High-mass X-ray binaries (HMXBs) are a class of binary systems
composed of a normal star of spectral type O or B
and a degenerate star (either a neutron star or a black hole).
The strong X-ray emission is caused by
the accretion of a fraction of the stellar wind ejected by the donor
star or through mass transfer via Roche-lobe
overflow (see e.g. \citealt{Treves88} and references therein).
There are about 250 known HMXBs in our Galaxy and in the Magellanic Clouds
(\citealt{Liu05}; \citealt{Liu06}; \citealt{Krivonos12}),
observed over a wide range of luminosities, from $\sim 10^{32}$~erg~s$^{-1}$
to $\sim 10^{39}$~erg~s$^{-1}$.
The members of this class are historically divided into two sub-classes:
HMXBs with OB supergiants (hereafter supergiant X-ray binaries, SGXBs)
and HMXBs with a rapidly rotating OBe main-sequence or giant star
accompanied by a transient circumstellar disc. 

Since the discovery of XTE~J1739$-$302 in 1998 (\citealt{Smith98};
see also \citealt{Sunyaev03})
and the identification of its optical counterpart \citep{Smith03}, 
a new generation of satellites observing the X-ray sky
led to the discovery of SGXBs with peculiar properties,
called supergiant fast X-ray transients 
(SFXTs; \citealt{Smith04}; \citealt{Sguera05}; \citealt{Negueruela06}),
which are the subject of this paper.
In contrast to the classical SGXBs, which are mostly bright
persistent sources showing moderate variability,
SFXTs show bright ($\sim 10^{36}-10^{37}$~erg~s$^{-1}$) and short
flares (lasting from minutes to hours).
The dynamic range of the flux change is three to five orders of magnitude,
from a quiescent luminosity of $\sim 10^{32}-10^{33}$~erg~s$^{-1}$ up to the 
outburst peak luminosity (\citealt{intZand05}; \citealt{Walter07}).
The \sw\ monitoring of a sample of five SFXTs (\citealt{Romano07}; 
\citealt{Sidoli08}; \citealt{Romano08})
showed that the short flares are the brightest part
of a much longer outburst event that typically lasts a few days.
In periods without outbursts, SFXTs show a fainter flaring activity
with luminosities of $\sim 10^{33}-10^{34}$~erg~s$^{-1}$.
The X-ray spectra during outbursts can be fitted with models that are commonly
used to describe the X-ray emission from HMXBs hosting neutron stars,
such as an absorbed power-law with cut-offs at $\sim 15-30$~keV (e.g. \citealt{Walter07}).

Although many SFXTs are in general remarkably different from the classical SGXBs,
some of them, the so-called \emph{intermediate SFXTs}, have properties
similar to some SGXBs, such as Vela~X$-$1 
(\citealt{Kreykenbohm08}; \citealt{Doroshenko11}),
and 4U~1907$-$09 \citep{Doroshenko12}.
\citet{Doroshenko11,Doroshenko12} attempted to link the dipping activity
observed in Vela~X$-$1 and 4U~1907$-$09 to the flaring activity
of SFXTs. They suggested that these two sources might
constitute a missing link between the SGXBs and SFXTs. 

\citet{Liu11} proposed that some of the SFXTs might have evolved from Be/X-ray binaries.
However, it is still not clear if SFXTs
come from an evolutionary path that is different from that of classical SGXBs.

Several accretion models have been proposed to explain
the properties of SFXTs. Nonetheless, a decisive explanation of
the peculiar transient behaviour of these sources is still
lacking. The proposed mechanisms are related
to the properties of either the supergiant wind or the
accreting neutron star. 
A first set of models involves accretion from a 
spherically symmetric inhomogeneous or anisotropic wind, 
where the flares are produced by the accretion of denser 
regions of matter expelled from the donor star (e.g. \citealt{intZand05};
\citealt{Negueruela08}; \citealt{Walter07}; \citealt{Sidoli07}; \citealt{Chaty13}).
The second set of models involves accretion-gating mechanisms, where the 
direct accretion onto a neutron star is halted by a magnetic or 
centrifugal barrier (depending on the spin period and the magnetic
field at the surface of the neutron star) and the high dynamic
range is caused by transitions through different regimes of accretion
(\citealt{Grebenev07}; \citealt{Bozzo08}; \citealt{Ducci10}; 
\citealt{Lutovinov13}; \citealt{Drave14} and references therein).
Other proposed scenarios associate the flaring activity with the formation
and dissipation of temporary accretion discs 
(\citealt{Smith06}; \citealt{Ducci10} and references therein).

Several studies on the expected number of Galactic HMXBs and their
X-ray luminosity functions have been made. 
These works tackle the problem either from observational 
(e.g. \citealt{Voss10}; \citealt{Lutovinov13}; \citealt{Doroshenko14})
or evolutionary points of view
(e.g. \citealt{Lipunov09}; \citealt{Belczynski08}; \citealt{Iben95}; \citealt{Dalton95};
\citealt{vandenHeuvel76}; \citealt{Meurs89}).
Except for the study of \citet{vandenHeuvel12}, 
who concluded that the number of discovered SFXTs
supports the number of HMXBs obtained from evolutionary models,
the expected number of SFXTs in the Galaxy 
has not been determined yet, nor have they been compared with the results
of these previous works.

The aim of this paper is to evaluate the expected number 
of SFXTs in the Galaxy to understand whether the SFXTs
are only a small sample of a very special case or rather a
significant subclass of HMXBs. The methods and input
information adopted and the results are shown in 
Sect. \ref{sect. Calculations}.
In Sect. \ref{sect. discussion} the results
and a comparison between the expected number of Galactic SFXTs calculated
in this work and the expected number of Galactic HMXBs
derived in other studies are discussed.

\section{Calculations}
\label{sect. Calculations}

We calculated the total number of SFXTs in the Galaxy adopting
two different approaches:
\begin{itemize}
\item a semi-analytical method, where the outbursts of SFXTs
      are treated as events occurring in a given interval of time
      with a known average rate that is independent of the time
      of the observation;
\item a series of Monte Carlo simulations of a population of HMXBs with the
      flaring properties of known SFXTs. This part of the work is
      based on the novel method of modelling XRB populations 
      of \citet{Doroshenko14}.
\end{itemize}
For the calculation we relied on two data sets:
the \emph{100-month \sw/BAT catalogue} (\citealt{Romano14}, hereafter R14; 
see Sect. \ref{sect. sw bat}) and the cumulative luminosity distributions
and duty cycles of SFXTs obtained by \citet[][hereafter P14]{Paizis14}
from a systematic analysis of about nine years of \inte\ data (17$-$100~keV).

The two datasets are based on all-sky hard X-ray surveys.
The \emph{Burst Alert Telescope} (BAT; \citealt{Barthelmy05})
on board the NASA satellite \sw\ \citep{Gehrels04} is a coded mask
telescope with a field of view of $\sim 2.29$~sr down to 5\% coding fraction
(percentage of the detector array shadowed by the mask).
It operates over the $15-150$~keV energy band \citep{Krimm13}.
The \emph{Imager on Board INTEGRAL Satellite} (IBIS; \citealt{Ubertini03})
is one of the telescopes on board the ESA satellite 
\emph{INTErnational Gamma-Ray Astrophysics Laboratory} (INTEGRAL; \citealt{Winkler03}).
It is a coded mask telescope with a field of view of $29^\circ\times29^\circ$
and is composed of two detectors: the \emph{INTEGRAL Soft Gamma-Ray Imager} 
(ISGRI; \citealt{Lebrun03}) which operates in $15-400$~keV band, and the
\emph{Pixellated Imaging Coesium Iodide Telescope} 
(PICsIT; \citealt{Dicocco03}) which operates in the $180-2000$~keV band. 

The wide fields of view and high sensitivities of these two instruments
are very useful for serendipitous surveys. Therefore,
they frequently observed SFXTs in outburst.
In particular, IBIS/ISGRI has a better short-term ($\sim$2~ks) sensitivity than BAT.
On the other hand, BAT provides a more uniform exposure over the Galactic plane
thanks to the wider field of view.

Despite the favourable characteristics provided by these and other X-ray satellites
for this type of studies, the detection of the entire population is difficult
because of the peculiar transient properties of SFXTs.
Therefore, a part of the population of Galactic SFXTs (especially those
with lower outburst rates and larger distances) 
has most likely not been detected yet.

\subsection{\sw/BAT}
\label{sect. sw bat}

The total number of SFXTs in the Galaxy can be derived
statistically from the number of the observed SFXTs and 
their outburst rates. Our calculation is based on the method
developed by \citet{Muno08} to constrain the number of Galactic
magnetars. As input data we used the detections in excess of
$5\sigma$ (corresponding to a $15-50$~keV limiting flux of 
$F_{\rm min} \approx 8.24 \times 10^{-10}$~erg~cm$^{-2}$~s$^{-1}$) from R14.
An outburst event produced by a SFXT typically contains several
flares that can lead to several detections per day. Since our
calculation requires the number of observed outbursts for any
given SFXT, if more than one detection occurred in a day we
considered only the detection with the highest flux.
In Table \ref{table1} the number of BAT orbital-averaged 
detections of outbursts $N_{\rm BAT}$ and the outburst rate $r=N_{\rm BAT}/T$, where
$T = 287.7$~d is the average monitoring period, are reported
for all the SFXTs detected with BAT.

The cumulative luminosity distribution of the observed flares
of SFXTs varies from source to source (see P14 and R14). 
For each SFXT, the lowest detected
luminosity depends on the sensitivity of the instrument
and the distance of the source from the observer.
Therefore, relatively distant SFXTs with intrinsically low 
outburst rate have most likely not been detected at all. An example is 
IGR~J11215$-$5952, a peculiar SFXT at $\approx 7.3$~kpc 
\citep{Coleiro13} which shows periodic bright flares 
($P_{\rm orb}\approx 165$~d; outburst rate $r=0.006$~d$^{-1}$; 
\citealt{Romano09}). 
It was never detected by BAT because of its 
relatively low $15-50$~keV flux during the outbursts 
($\approx 2-12 \times 10^{-10}$~erg~cm$^{-2}$~s$^{-1}$; 
\citealt{Lubinski05}; \citealt{Sidoli06}).
The first step of our calculation was therefore to estimate the fraction
of the Galaxy completely surveyed by BAT for SFXTs
that have properties of the observed SFXT sample.
The prototypical\footnote{Prototypical SFXTs have a dynamic range exceeding
$10^3$ (see e.g. P14). In addition to SAX~J1818.6$-$1703 
the other prototypical SFXTs are XTE~J1739$-$302, IGR~J17544$-$2619, and IGR~J08408$-$4503.} 
SFXT SAX~J1818.6$-$1703 shows the lowest luminosities among all the
SFXTs detected by BAT. Therefore, 
we calculated the largest distance that an SFXT with the same properties
as SAX~J1818.6$-$1703 can have to still be detected by BAT with a confidence
level of $~90$\%. According to Poisson statistics, the probability
that any given SFXT is detected is $p=1 - e^{-N}$.
Assuming $p=0.9$, we find $N\approx 2.3$. 
Since $N$ is a discrete variable, for the following calculation of $f$
we approximated $N$ as $N \sim 2$. 
From the BAT luminosity distribution of SAX~J1818.6$-$1703
we find that the luminosities corresponding to the two detections with highest fluxes
are higher than $L_{\rm min} \approx 1.4 \times 10^{36}$~erg~s$^{-1}$ ($15-50$~keV).
Flares with these luminosities can be observed by BAT up to a distance
of $R_{\rm loc} = \sqrt{L_{\rm min}/(4 \pi F_{\rm min})} \approx 3.8$~kpc.
Using the model for stellar mass distribution of the Galaxy
of \citet{Cordes02}, $R_{\rm loc}$ derived above, and the BAT sky coverage 
of 80\%$-$94\%~day$^{-1}$ \citep{Krimm13}, we found that the fraction 
of stellar mass in the spiral arms surveyed by BAT for SFXTs each day 
is $f=11\%-13\%$.
Taking into account the position of the SFXTs detected by BAT 
in the Galactic plane, $s=5 \pm 1$ of them are surveyed each day by BAT.

 \begin{table}
 \begin{center}
 \caption{Orbitally averaged detections ($>5\sigma$) throughout the \sw\ mission (2005-02-12 to 2013-05-31) 
          from R14. When more than one detection was made on a given day, 
          only one detection has been counted.}
\label{table1}
\begin{tabular}{lcccc}
 \hline
 \hline
 \noalign{\smallskip}
Name               &\multicolumn{2}{c}{distance}&$N_{\rm BAT}$&     r     \\
                   &   (kpc)      &     Ref.    &$>5\sigma$& (d$^{-1}$) \\
                   &              &             &orbital   &           \\
  \noalign{\smallskip}
 \hline
 \noalign{\smallskip}
IGR~J08408$-$4503  & $3.4\pm0.3$ &     (1)      &    5      &  0.017    \\
IGR~J16328$-$4726  &   $3-10$    &     (2)      &    3      &  0.010    \\
IGR~J16418$-$4532  &    $13$     &     (3)      &   16      &  0.054    \\
IGR~J16465$-$4507  & $12.7\pm1.3$&     (1)      &    1      &  0.003    \\
IGR~J16479$-$4514  &    $4.9$    &     (4)      &   61      &  0.208    \\
XTE~J1739$-$302    &    $2.7$    &     (4)      &   29      &  0.102    \\
IGR~J17544$-$2619  &    $3.6$    &     (4)      &   23      &  0.083    \\
SAX~J1818.6$-$1703 &    $2.1$    &     (5)      &   17      &  0.062    \\
AX~J1841.0$-$0536  & $7.8\pm0.7$ &     (1)      &   16      &  0.055    \\
AX~J1845.0$-$0433  &    $3.6$    &     (6)      &    8      &  0.028    \\
IGR~J18483$-$0311  &    $2.8$    &     (5)      &   24      &  0.083    \\
  \noalign{\smallskip} 
  \hline
  \end{tabular}
  \end{center}
Notes: (1) \citet{Coleiro13}; (2) \citet{Fiocchi13}; (3) \citet{Chaty08}; 
(4) \citet{Rahoui08}; (5) \citet{Torrejon10}; (6) \citet{Coe96}.
\end{table}

The probability that an SFXT in the region of the Galaxy
surveyed by BAT is detected during an outburst is $1 -e^{-rT}$,
where $r$ is the outburst rate reported in Table \ref{table1}.
According to the binomial distribution, given $n \geq s$ SFXTs
in the surveyed region and $N \geq n$ SFXTs in the Galaxy,
the probability that $s$ SFXTs are detected in outburst is:
\begin{equation} \label{eq. prob.}
p(s|rT,f,N) = \sum\limits_{n=s}^N \left [ (1 -e^{-rT})^s (e^{-rT})^{n-s}  \binom{n}{s} \times  f^n(1-f)^{N-n} \binom{N}{n} \right ]
\end{equation}
where the first part of the right-hand side of Eq. (\ref{eq. prob.})
is the probability that $s$ out of $n$ SFXTs in the surveyed region
are detected and the second part is the probability that $n$ out of $N$
SFXTs are located in a fraction $f$ of the Galaxy 

According to the Bayes theorem, the conditional probability 
$p(s|rT,f,N)$ defined in Eq. (\ref{eq. prob.})
can be inverted to obtain $p(N|rT,f,s)$,
namely the probability that there are $N$ SFXTs
given that $s$ are found in a fraction $f$ of the Galaxy.
In our calculations, we assumed a uniform prior distribution
(for a similar application of this method, on magnetars, see \citealt{Muno08}).

Given the average outburst rate $\bar{r}=0.06$~d$^{-1}$ (see Table \ref{table1})
of the SFXTs within $R_{\rm loc}$
and assuming $s=5 \pm 1$, $T=287.7$~d, and $f$ ranging from 11\% to 13\%,
we obtained that the total number of SFXTs in the Galaxy 
for the given average outburst rate is 
$N_{\rm \bar{r}}=37{+53 \atop -22}$ (uncertainties at 90\% confidence level).

\begin{figure}
\begin{center}
\includegraphics[width=\columnwidth]{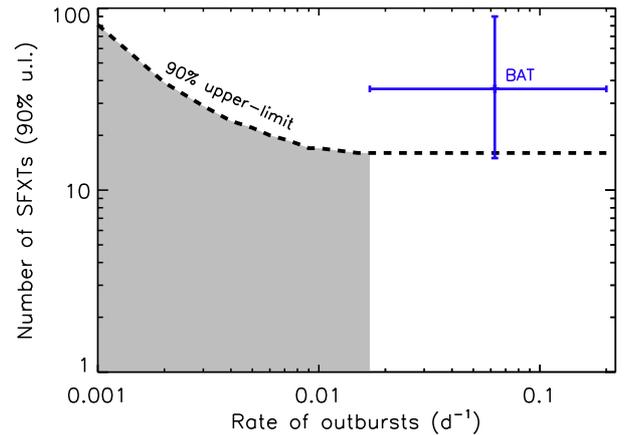}
\end{center}
\caption{90\% upper limits of the total number of SFXTs in the Galaxy as a function of the outburst rate.
         The dashed line corresponds to the 90\% upper limits
         of $N$ obtained from the probability distribution $p(N|rT,f,s)$.
         The grey area shows for each assumed outburst rate, 
         the allowed $N$, given $s=0$, $T=287.7$~d, and $f$ ranging from $11\%$ to $13\%$.
         The value of $N_{\rm \bar{r}}$ obtained assuming $\bar{r}=0.06$~d$^{-1}$,
         $s=5 \pm 1$, $T=287.7$~d, and $f=11\%-13\%$ is also plotted.}
\label{figure1}
\end{figure}

The SFXTs used for this calculation have outburst rates ranging 
from $\approx0.017$~d$^{-1}$ to $\approx0.2$~d$^{-1}$.
However, we know that there are other SFXTs and candidate SFXTs with lower outburst rates.
In addition to IGR~J11215$-$5952 with $r=0.006$~d$^{-1}$,
\inte\ detected other SFXTs candidates:
IGR~J20188$+$3647, IGR~J18159$-$3353, IGR~J21117$+$3427 (\citealt{Sguera06} and \citealt{Bodaghee07}),
IGR~J17541$-$2252 (\citealt{Turler06} and \citealt{Bodaghee07}),
IGR~J10043$-$8702, IGR~J10500$-$6410, IGR~J15283$-$4443 \citep{Paizis06},
IGR~J18462$-$0223 \citep{Grebenev07b}.
They have been detected only once in $\sim$11 years (except for IGR~J18462$-$0223), 
therefore their outburst rates are likely lower than 0.017~d$^{-1}$.
Driven by the high number of SFXTs and candidate SFXTs with $r\lesssim0.02$~d$^{-1}$,
we calculated the 90\% complete upper limits of the number of Galactic SFXTs
with outburst rates in the range $0.001\leq r \leq 0.017$~d$^{-1}$.
In Fig. \ref{figure1} we plot the upper-limits of the expected number
of Galactic SFXTs as a function of the outburst rate (dashed line).
We derived the upper limits from the probability distribution $p(N|rT,f,s)$
that can be obtained by inverting Eq. (\ref{eq. prob.})
and assuming $s=0$ (because BAT did not detect any SFXT
with $r < 0.017$~d$^{-1}$ within $R_{\rm loc}$), $T=287.7$~d, and $f=11-13\%$.
The grey region in Fig. \ref{figure1} shows the solutions allowed by the probability distribution
$p(N|rT,f,s)$ for $s=0$ and $0.001\leq r \leq 0.017$~d$^{-1}$.
The value of $N_{\rm \bar{r}}$ obtained above is also plotted.

\subsection{INTEGRAL}
\label{sect. integral}

A large portion of all observed SFXTs has been discovered by \emph{INTEGRAL}.
The reason is that while the cumulative sensitivies of \emph{Swift}/BAT and
\emph{INTEGRAL}/ISGRI surveys are similar, the latter has a sensitivity 
that is a factor of four better 
on the short time-scales relevant for detecting SFXT flares. 
Therefore, we also estimated the number of the missing SFXTs using the
\emph{INTEGRAL} data. Unfortunately, the \emph{INTEGRAL}/ISGRI exposure is
distributed much less uniformly across the sky than for
\emph{Swift}/BAT, which has to be taken into account. This is not trivial with
the simple approach presented above, therefore we used a series of Monte Carlo
simulations instead. This also allowed us to independently verify the results
obtained above with a different method.

We assumed that SFXTs have the same spatial distribution as classical HMXBs
\citep{Grimm02,Coleiro13,Doroshenko14} and the same flaring properties as the sample
of known SFXTs, which have been investigated recently by P14.
Based on the assumed spatial distribution, we assigned a
random position within the Milky Way for each simulated source 
and calculated the survey exposure in the
respective direction. We assumed that SFXTs can only be detected in flaring
state, and the effective flare exposure was calculated by multiplying the total
exposure with the duty cycle randomly chosen among the values derived for known
SFXTs (see Table~3 of P14). P14 assumed an average
flare duration corresponding to a science window (ScW) length,
hence the number of flares expected to occur during the
survey from a given SFXT equals the number of ScWs of Table~2 
of P14. For each flare we assigned a luminosity 
randomly sampled from one of the cumulative flare
luminosity distribution functions reported by P14 (randomly chosen
for each source). Distances to the simulated sources are known, which makes it
trivial to calculate the flare flux distribution and highest flux for each source.
The sensitivity of ISGRI on an ScW timescale is
$\sim2\times10^{-10}$\,erg\,cm$^{-2}$\,s$^{-1}$ (17$-$60~keV), 
therefore we assume that any flares with a flux exceeding this limit 
will be detected. As an estimate of the total
ISGRI exposure we took the exposure maps from the $\sim$nine-year
Galactic plane survey \citep{Krivonos12}. 

We applied a similar approach
to \emph{Swift}/BAT, where we assumed slightly lower sensitivity to flares of
$8\times10^{-10}$\,erg\,cm$^{-2}$\,s$^{-1}$, and a uniform exposure of 287.7\,d. The main
difference with the simpler estimation presented above is that we imposed no
restriction on distance to the known sources.

We found that both \inte\ and \sw\ are expected to detect at least once 46(2)\% of the SFXTs in Galaxy.
Of the known confirmed SFXTs, IGR~J08408$-$4503 has the lowest number of ScWs, three,
where the source is detected by \inte.
If we take three as the minimum number of detections required to consider a source
as an SFXT (i.e. the brightest flares must be detected), 
\inte\ is expected to detect $\sim$27\% of SFXTs in the Galaxy.
There are 14 confirmed SFXTs\footnote{In addition to the SFXTs listed in Table \ref{table1},
the other three SFXTs are IGR~J11215$-$5952, IGR~J16195$-$4945, and IGR~J17354$-$3255.}
and eight candidates (seven of which are observed only once), 
which makes the total expected number
of SFXTs in the Milky Way between 29 and 52.

\section{Discussion}
\label{sect. discussion}

We applied two methods to \sw/BAT and \inte/ISGRI data to derive the total number of 
Galactic SFXTs. The numbers obtained with these methods agree.
We found that the total number
of SFXTs in the Galaxy is $\approx 37{+53\atop-22}$.
We note that this estimate strongly depends on the luminosity distribution of the flares,
which is reasonably well constrained, however.
Since \emph{INTEGRAL} is more likely to
detect brighter flares and the portion of sources detectable with \emph{INTEGRAL}
and \emph{Swift} decreases with average flare luminosity, there
may be more SFXTs in Galaxy. On the other hand, there are probably
fewer than three hundred HMXBs in the Milky Way \citep{Doroshenko14}, 
which makes our result of 15$-$90 SFXTs seem reasonable.

The fraction of SFXTs in SGXBs is fundamentally related to their evolutionary origin,
therefore it would be interesting to estimate it observationally and compare it
with theoretical predictions for various models.
As mentioned in Sect. \ref{sect. intro}, 
the peculiar properties of SFXTs are commonly
associated either with wind properties or with inhibition 
of the accretion by the neutron star centrifugal 
and magnetic barriers (\citealt{Bozzo08}; see also \citealt{Chaty13} for a review).
In the latter case, it is easy to estimate the expected 
fraction of SFXTs among the SGXBs.
\citet{Meurs89}, \citet{Dalton95}, and \citet{Iben95} calculated the expected number 
of massive X-ray binaries in the Galaxy. 
In particular, \citet{Dalton95} modelled the HMXB population in the Galaxy 
for different evolutionary scenarios 
(initial mass function slope $\Gamma$, birthrate, binary fraction,
distribution of binary separations and binary mass ratios, and the fraction $f_{\dot{M}}$ of the matter
lost from a binary system).
The number of HMXBs with $L_x\gtrsim 10^{35}$~erg~s$^{-1}$ 
predicted by \citet{Dalton95} ranges between about 200 and 400.
In the framework of the results obtained by \citet{Lutovinov13},
who reported that the transient behaviour of SFXTs is caused by an inhibition
of accretion onto the neutron star, it follows that 
all the observed SFXTs would be persistent with a luminosity
$L_x\gtrsim 10^{35}$~erg~s$^{-1}$ rather than highly variable
with an observed average luminosity $\ll$10$^{35}$~erg~s$^{-1}$.
It is therefore of particular interest to compare the number 
of observed and predicted SFXTs $+$ SGXBs with the estimate 
provided by the evolutionary model of \citet{Dalton95}
in the framework of the hypothesis of \citet{Lutovinov13}.
There are 14 confirmed and 8 candidate observed SFXTs and
18 SGXBs observed with $L_x\gtrsim 10^{35}$~erg~s$^{-1}$ \citep{Lutovinov13}.
From the X-ray luminosity function of \citet[][see figure 2, right panel]{Doroshenko14} 
there are 80$-$100 HMXBs with $L_x\gtrsim 10^{35}$~erg~s$^{-1}$.
We point out that this number does not take into account SFXTs.
Indeed, they have observed average luminosities $\ll 10^{35}$~erg~s$^{-1}$.
Therefore, there are $\sim 32$ (40 including candidate SFXTs)
observed SFXTs$+$SGXBs in total and $\sim100-200$ predicted HMXBs 
(80$-$100 HMXBs with $L_x\gtrsim 10^{35}$~erg~s$^{-1}$ plus 15$-$90 SFXTs),
which are consistent with the evolutionary constraints obtained by \citet{Dalton95}
for $L_x\gtrsim 10^{35}$~erg~s$^{-1}$.

\section{Conclusions}

We derived for the first time
the expected number of SFXTs in Galaxy. We applied two different
approaches and considered two sets of data, the 100-month \sw/BAT
catalogue (R14) and the results presented by P14 who performed
a systematic analysis of $\sim$nine-years of \inte\ data.
The total number of SFXTs in the Galaxy is $\approx 37{+53\atop-22}$.
This value agrees with the total number of Galactic HMXBs
with $L_{\rm x} \geq 10^{33}$~erg~s$^{-1}$ \citep{Doroshenko14}
and with the number of expected HMXBs in the Galaxy obtained
from studies of the evolution of high-mass binary systems,
and shows that SFXTs represent a significant fraction of all HMXBs.
In the framework of the scenario proposed by \citet{Lutovinov13},
the number of Galactic SFXTs derived in this work agrees with
the expected number of HMXBs with $L_{\rm x} \geq 10^{35}$~erg~s$^{-1}$
obtained by \citet{Dalton95} from evolutionary arguments (which does not
take into account the effects of gating mechanisms on the observed X-ray
luminosity).

Observing SFXTs in nearby galaxies is greatly complicated by their
very fast transient behaviour, which makes them difficult to detect
in outburst with the typical observation lengths of \emph{Chandra} and
\emph{XMM-Newton}.
Indeed, the first and only extragalactic SFXTs candidate has been only recently claimed to be observed by \citet{Laycock14}.
The detection of SFXTs and an estimate of their number in the Magellanic Clouds
would be important for understanding their dependence on the metallicity of a galaxy.
Indeed, metallicity plays an important role in the radiatively driven wind mechanisms
of massive stars and consequently in the values of mass-loss rate and 
velocity of the winds of donor stars of HMXBs \citep{Lamers99}.
Since the Large and Small Magellanic Clouds have a metallicity of about
40\% and 10\% of that of the Milky Way, respectively, they could represent
an ideal target for this type of studies.

\begin{acknowledgements}
We thank the referee Sylvain Chaty for his useful comments,
which helped to improve the paper.
This work is partially supported by the Bundesministerium f\"ur
Wirtschaft und Technologie through the Deutsches Zentrum f\"ur Luft
und Raumfahrt (grant FKZ 50 OG 1301).
VD and AS thank the Deutsches Zentrum f\"ur Luft- und Raumfahrt (DLR) and
Deutsche Forschungsgemeinschaft (DFG) for financial support
(grant DLR 50 OR 0702).
MS acknowledges support by the Deutsche Forschungsgemeinschaft
through the Emmy Noether Research Grant SA 2131/1-1.
PR acknowledges contract ASI-INAF I/004/11/0.
We thank H.A. Krimm and C.B. Markwardt  for helpful discussions. 
This research has made use of the IGR Sources page maintained 
by J. Rodriguez \& A. Bodaghee (\url{http://irfu.cea.fr/Sap/IGR-Sources/}).
\end{acknowledgements}

\bibliographystyle{aa} 
\bibliography{num_sfxts}

\end{document}